\documentclass[11pt]{article}
\usepackage{epsfig,dsfont}

\begin{document}  
\sffamily

\thispagestyle{empty}
\vspace*{15mm}

\begin{center}

{\LARGE 
No coincidence of center percolation and 
\vskip2mm
deconfinement in SU(4) lattice gauge theory}
\vskip25mm
Michael Dirnberger$^a$, Christof Gattringer$^a$ and Axel Maas$^b$
\vskip10mm
$^a$Institute for Physics, Karl-Franzens University Graz, \\
Universit\"atsplatz 5, 8010 Graz, Austria 
\vskip5mm
$^b$Institute for Theoretical Physics, Friedrich-Schiller-University Jena \\
Max-Wien Platz 1, D-07743 Jena, Germany

\end{center}
\vskip30mm

\begin{abstract}
We study the behavior of center sectors in pure SU($4$) lattice gauge theory at finite
temperature. The center sectors are defined as spatial clusters of neighboring sites
with values of their local Polyakov loops near the same center elements. We study
the connectedness and percolation properties of the center clusters across the
deconfinement transition. We show that for SU($4$) gauge theory deconfinement
cannot be described as a percolation transition of center clusters, a finding which is
different from pure SU($2$) or pure SU($3$) Yang Mills theory, where the percolation 
description even allows for a continuum limit.
\end{abstract}

\newpage
\setcounter{page}{1}

\noindent
{\Large Introduction and outline}
\vskip3mm

\noindent
Understanding the high temperature transition of QCD to a phase of deconfined
quarks and gluons is still an open problem. Finding a suitable description 
of the transition and the plasma phase is essential for understanding current and
upcoming results from heavy ion experiments. An interesting approach is the idea
of describing the deconfinement transition as a percolation 
phenomenon and to explore its phenomenological consequences. 

For the simpler case of pure gauge theory the idea that 
percolation of clusters related to the center of the gauge group can
be used to describe deconfinement goes back to \cite{satz} for the case of SU($2$)
and to \cite{fortunato} for SU($N$) with $N > 2$. The clusters were constructed from
the local Polyakov loops, i.e., they reflect the properties of static
quark sources under the transformation with the center group $\mathds{Z}_N$ of
SU($N$). More recently it was argued that
for the cases of SU($2$) and SU($3$) even a continuum limit of the percolation
description is possible for suitably defined clusters. 
First results for full lattice QCD were presented in \cite{borsanyi}. 

With the encouraging results available for SU($2$) and SU($3$) one may ask the
question whether the percolation description of deconfinement  based on the
center group is suitable for all gauge groups SU($N$).  This is not a priori
clear: The number of center elements is $N$, such that if the center sectors are
occupied uniformly the probability that a site belongs to a particular sector is
$1/N$. For $N > 3$, this probability $1/N$ is below the critical occupation
probability $p_c \sim 0.316$ of random percolation on a
three-dimensional simple cubic lattice. This implies that if the percolation picture of
deconfinement were to apply also for SU($N$) with $N > 3$, the deconfinement
transition must be accompanied by the onset of very strong correlations between
the center phases of neighboring sites. The main goal of the current letter 
(see also \cite{michael}) is to
establish or disproof the existence of such strong correlations and the
corresponding percolation picture for SU($N$) Yang-Mills theory at $N > 3$. 

It should be kept in mind that rigorous results for a complete
description of thermal transitions, i.e., the same transition temperatures 
of the percolation and the thermal transition and matching critical exponents,
are only available for continuous transitions. For first order transitions only 
numerical simulations or results for simple models suggest that percolation 
may be used to effectively describe thermal transitions in some cases (see, e.g.,
\cite{first1,first2}). The current paper provides a counter example.

\newpage
\noindent
{\Large Setting of the calculation}
\vskip3mm

\noindent
We work with pure SU($4$) lattice gauge theory using Wilson's formulation. The fundamental
degrees of freedom are the SU($4$) valued link variables $U_\mu(x)$ with $\mu = 1,2,3,4$.
$x$ denotes the sites of a $N_s^{\,3} \times N_t$ lattice with periodic boundary conditions. 
$N_t$ is the temporal extent of the lattice and $1/N_t$ is the temperature in lattice
units. The action is the usual sum over all plaquettes and the partition sum is
obtained by integrating over all gauge configurations. We work on lattices of sizes
$20^3 \times 6$ to $40^3 \times 10$ and use ensembles of typically 500 configurations. 
The update was done using heat bath and overrelaxation steps \cite{maas}, and the error bars we show 
are statistical errors from a single elimination
Jackknife analysis corrected for autocorrelation. 
For the scale (from the string tension) and the
deconfinement temperature we use the values from \cite{scale}.  

The basic observable in our study is the traced local Polyakov loop $L(\vec{x})$,  
\begin{equation}
L(\vec{x}) \, = \, 
\mbox{Tr} \prod_{t=1}^{N_t} U_4(\vec{x},t) \; , 
\label{Ploopdef}
\end{equation}
i.e., the ordered product of all temporal gauge links at a spatial lattice point 
$\vec{x}$ and $\mbox{Tr}$ denotes the trace over color indices. 
$L(\vec{x})$ is a gauge invariant object that corresponds to a gauge transporter 
which closes around compactified time, interpreted as a static source of 
color flux. We will also consider the normalized
spatial average of the local Polyakov loops
which we denote by $P$,
\begin{equation}
P \; =  \; \frac{1}{V} \, \sum_{\vec{x}} L(\vec{x})\; ,
\label{averloop}
\end{equation}
where $V$ is the spatial lattice volume.  Due to translational
invariance $P$ and $L(\vec{x})$ have the same vacuum expectation value. 

\begin{figure}[t]
\begin{center}
\hspace*{-5mm}
\includegraphics[width=10cm,clip]{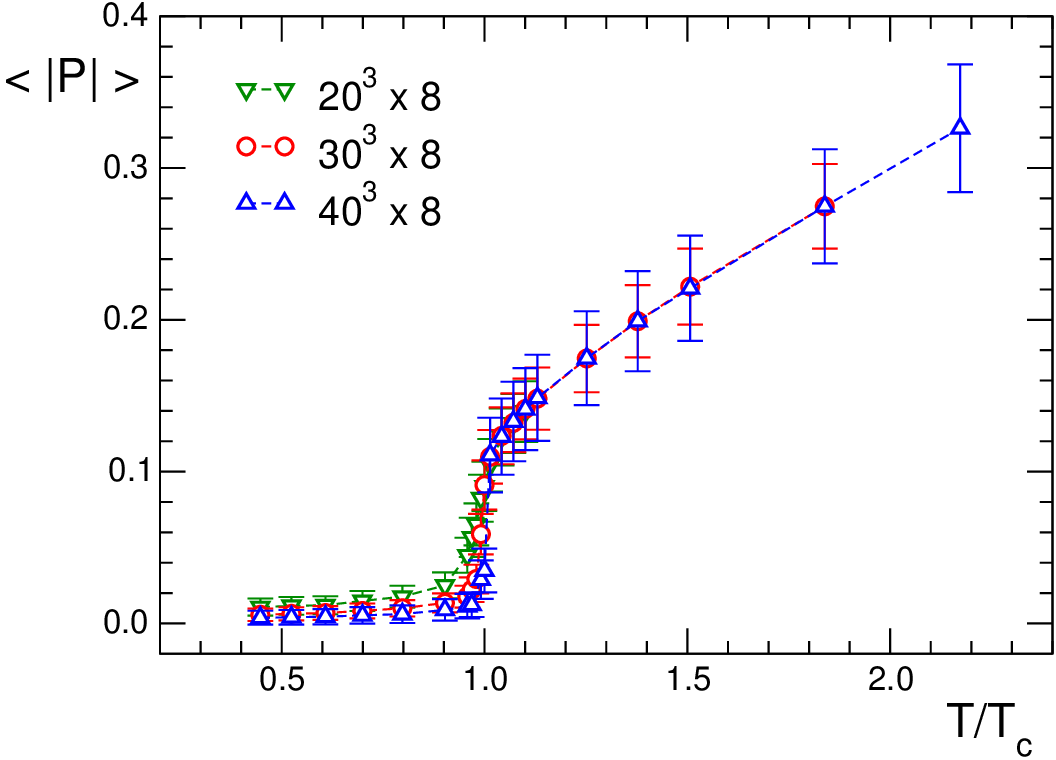} 
\end{center}
\vspace{-5mm}
\caption{
Expectation value $\langle | P | \rangle$ of the absolute value of the
Polyakov loop as a function of the temperature. We compare the results from lattice
sizes $20^3 \times 8$, $30 \times 8$ and $40 \times 8$ to illustrate the approach
towards the discontinuous first order transition. 
\label{ploopexp}}
\end{figure}

The Polyakov loop $L(\vec{x})$ (and also its spatial average $P$) transform
non-trivially under center transformations. For SU($4$) the center group is 
$\mathds{Z}_4 = \{1,i,-1,-i \}$ and in a center transformation all temporal links for
some fixed time slice $t = const$ are multiplied with a center element $z \in
\mathds{Z}_4$. While the action and the path integral are invariant under center
transformations, the local and averaged Polyakov loops transform as
\begin{equation}
L(\vec{x}) \; \longrightarrow \; z \, L(\vec{x}) \qquad \mbox{and} \qquad P \;
\longrightarrow \; z \, P \; .
\end{equation} 
Below the deconfinement temperature the vacuum is invariant under center
transformations and the expectation value of the Polyakov loops vanishes, i.e.,
$\langle L(\vec{x}) \rangle = \langle P \rangle = 0$. Above $T_c$ the center
symmetry is spontaneously broken which is signaled by  $\langle L(\vec{x})
\rangle = \langle P \rangle \neq 0$. In this spontaneous breaking of the center
symmetry, the phase of the expectation value spontaneously selects one  of the
four center values. The deconfinement transition of pure SU($4$) lattice gauge
theory  is of a pronounced first order. For later comparison in
Fig.~\ref{ploopexp} we show the  expectation value $\langle | P | \rangle$ as a
function of the temperature. 

The Polyakov loop corresponds to a static color source and its vacuum
expectation value is (after a suitable renormalization)  related to the free
energy $F_q$ of a single quark, $\langle L(\vec{x}) \rangle = \langle P
\rangle  \propto \exp(-F_q/T)$, where $T$ is the temperature  (the Boltzmann
constant is set to 1 in our units). Thus, when $\langle L(\vec{x}) \rangle = \langle P
\rangle$ vanishes $F_q$ is infinite and quarks are confined. On the other hand a
non-zero value $\langle L(\vec{x}) \rangle = \langle P
\rangle \neq 0$ implies finite $F_q$ and quarks are deconfined. Thus, for pure gauge
theory the deconfinement transition is linked to the spontaneous breaking of center
symmetry.

\vskip7mm
\noindent
{\Large Local Polyakov loops and cluster definition}

\begin{figure}[t!]
\begin{center}
\includegraphics[width=12.5cm,clip]{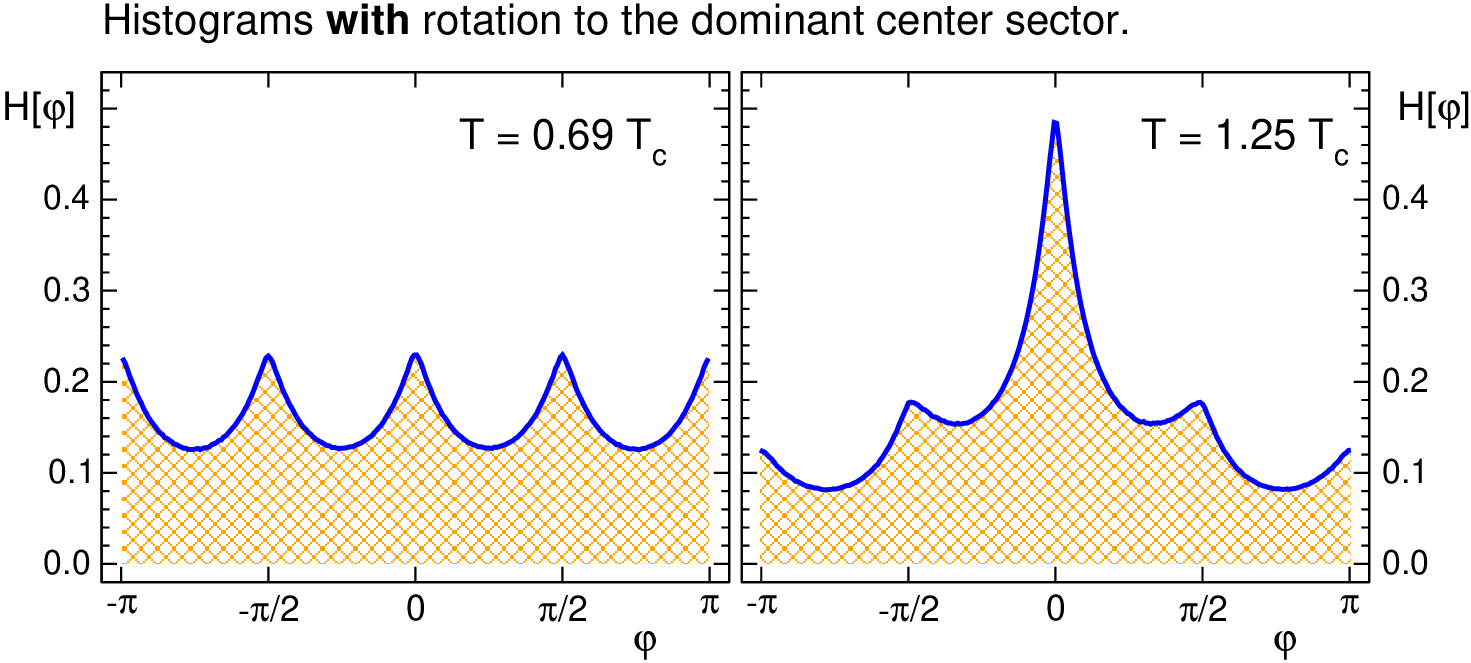}
\vskip5mm
\includegraphics[width=12.5cm,clip]{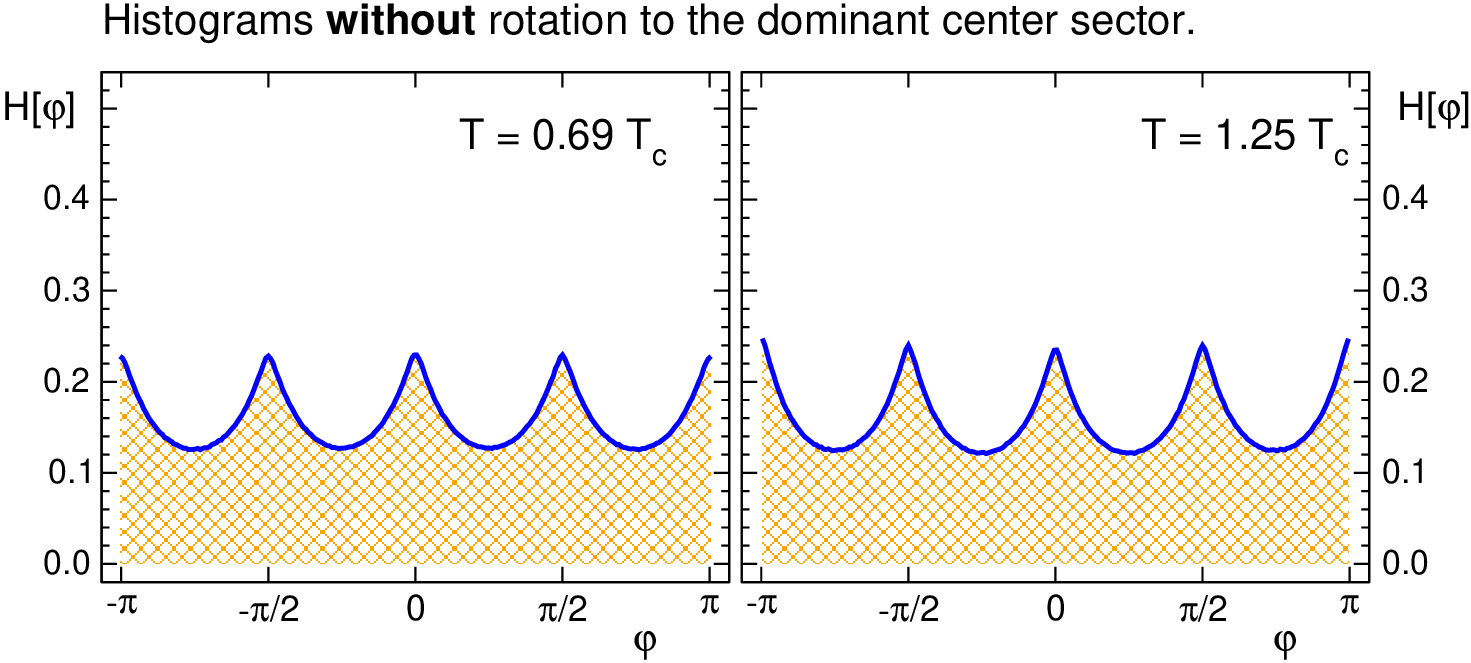}
\end{center}
\vspace{-5mm}
\caption{Histograms for the distribution of the phases $\varphi(\vec{x})$ of 
the local Polyakov loops $L(\vec{x})$. The results are from our $40^3 \times 6$ 
ensembles at temperatures $T=0.69 \, T_c$ (lhs.\ plots) and $T = 1.25 \,
T_c$ (rhs.). We compare the histograms with (top row of plots) and without
(bottom) center rotation bringing the dominant center sector to $\varphi \sim 0$. 
\label{phasehisto}}
\end{figure}

\vskip3mm
\noindent
So far we have only considered expectation values of the averaged  Polyakov loop $P$
without looking at the distribution of $L(\vec{x})$ at different  spatial lattice sites
$\vec{x}$. Now we study this local behavior. The local Polyakov loop $L(\vec{x})$ is a
complex number which we decompose into modulus and phase,  
\begin{equation}
L(\vec{x}) \; = \; \rho(\vec{x}) \, e^{ i \varphi(\vec{x}) } \; .
\end{equation}
While the distribution of the modulus is rather insensitive to the  temperature (it almost perfectly
follows the corresponding Haar measure distribution -- as for  SU($3$) \cite{gattringer}), the
distribution of the phase $\varphi(\vec{x})$  strongly depends on the temperature. In
Fig.~\ref{phasehisto} we show histograms for the values of the angles $\varphi(\vec{x})$. For the top
row of plots we applied center rotations of the individual configurations such that the dominant
sector is always the one with real and positive phase ($\varphi \sim 0$). In the bottom row of plots,
for comparison we show the distribution without rotating the dominant sector to $\varphi \sim 0$, and the 
random center rotation in our Monte Carlo  update averages over all four sectors.

The histograms for $T < T_c$ nicely illustrate the center symmetry: The phases
are distributed such that they have maxima near the center elements, which
correspond to phases $\varphi = 0, \pm \pi/2$ and $\varphi = \pi$ (or $-\pi$).
The maxima are of equal height and center symmetry is manifest. For $T > T_c$
in the plot with rotation of the dominant sector
(rhs.\ top plot), we see that one of the peaks ($\varphi = 0$) is much taller than the others
indicating that the corresponding center sector is considerably more populated.
The figure corresponds to the situation where the spontaneous symmetry breaking has selected the 
$\varphi = 0$ sector. Note that the spontaneous breaking is possible only in infinite volume, and
that our rotation of the dominant sector only mimicks this behavior. 
 It is interesting to note that also above $T_c$
the non-dominant center sectors are still populated, as there are local maxima
near $\varphi = \pm \pi/2, \pm \pi$. This indicates that above $T_c$ (at least
for not too high temperatures) there is still an admixture of local Polyakov
loops with a phase different from the dominant one.

Based on the phases of the local loops $\varphi(\vec{x})$
we now assign local sector numbers $n(\vec{x})$ to sites
$\vec{x}$ as follows:
\begin{eqnarray}
n(\vec{x}) = \;\;\;0 & \; \Leftrightarrow \; & \varphi(\vec{x}) \, \in \, [-\pi/4,\pi/4) \;, \\
n(\vec{x}) = \;\;\;1 & \; \Leftrightarrow \; & \varphi(\vec{x}) \, \in \, [\pi/4,3\pi/4) \;, \\
n(\vec{x}) = -1 & \; \Leftrightarrow \; & \varphi(\vec{x}) \, \in \, [-3\pi/4,-\pi/4) \;, \\
n(\vec{x}) = \;\;\; 2 & \; \Leftrightarrow \; & \varphi(\vec{x}) \geq  3\pi/4 \;\; \mbox{or} \;\;
\varphi(\vec{x}) < -3\pi/4 \; .
\end{eqnarray}
Using the local sector numbers $n(\vec{x})$ we can now define the center
clusters by putting two neighboring sites $\vec{x}$ and $\vec{y}$ into the same
cluster when $n(\vec{x}) = n(\vec{y})$. Usual cluster identification techniques
\cite{clusterfind} can then be used to find all the clusters on the lattice.

\vskip7mm
\noindent
{\Large Properties of center clusters}

\begin{figure}[t]
\begin{center}
\hspace*{-13mm}
\includegraphics[width=10.0cm,clip]{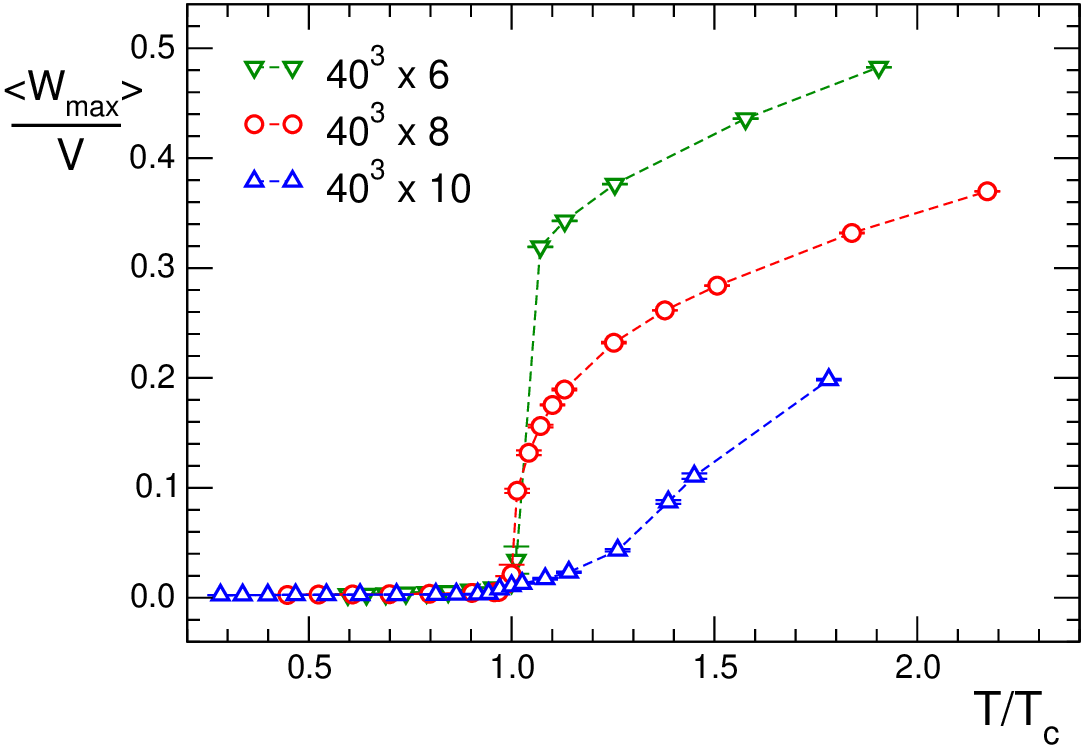}
\end{center}
\vspace{-5mm}
\caption{Weight of the largest cluster normalized by the spatial volume 
as a function of temperature.  
\label{weight}}
\end{figure}

\vskip3mm
\noindent
Having analyzed the distribution of the phases of the local Polyakov loops and
constructed the center clusters we can now study their properties as a function
of the temperature. The simplest quantity is the weight $W$ of a cluster which
is simply defined as the number of sites in the cluster. In particular by
$W_{max}$ we denote the weight of the largest cluster. In Fig.~\ref{weight} we
show the behavior of the expectation value $\langle W_{max} \rangle/V$  of the
weight of the  largest cluster normalized by the spatial volume as a function of
the temperature. For fixed $N_s = 40$  we compare the results from calculations
at three different values of $N_t$. As $N_t$ increases, the lattice spacing $a$
has to be decreased in order to stay at a fixed value of the temperature  $T = (a
N_t)^{-1}$. Thus as one increases $N_t$ at fixed $T$ the continuum limit is
approached, and comparing the curves for $\langle W_{max} \rangle$ at different 
$N_t$ allows one to study their behavior in the continuum limit.     

Below $T_c$ the three curves for $\langle W_{max} \rangle/V$ in Fig.~\ref{weight} 
fall on top of each other. For all three values of $N_t$ the expectation value 
$\langle W_{max} \rangle$ is small compared to the volume $V$ such that the ratio is close to 0.
At $T_c$ the largest cluster starts to grow quickly and, e.g., for the $N_t = 6$ ensembles
reaches at 2 $T_c$ a weight which is already half of the volume. However, it is obvious that the
growth rate above $T_c$ strongly depends on the temporal extent $N_t$. For larger values of
$N_t$, i.e., closer to the continuum limit, the growth rate is considerably slower, which is a
first hint that in the continuum limit $a \rightarrow 0$ ($N_t \rightarrow \infty$) the clusters
might not grow fast enough to give rise to percolation.

\begin{figure}[t]
\begin{center}
\hspace*{-5mm}
\includegraphics[width=9.3cm,clip]{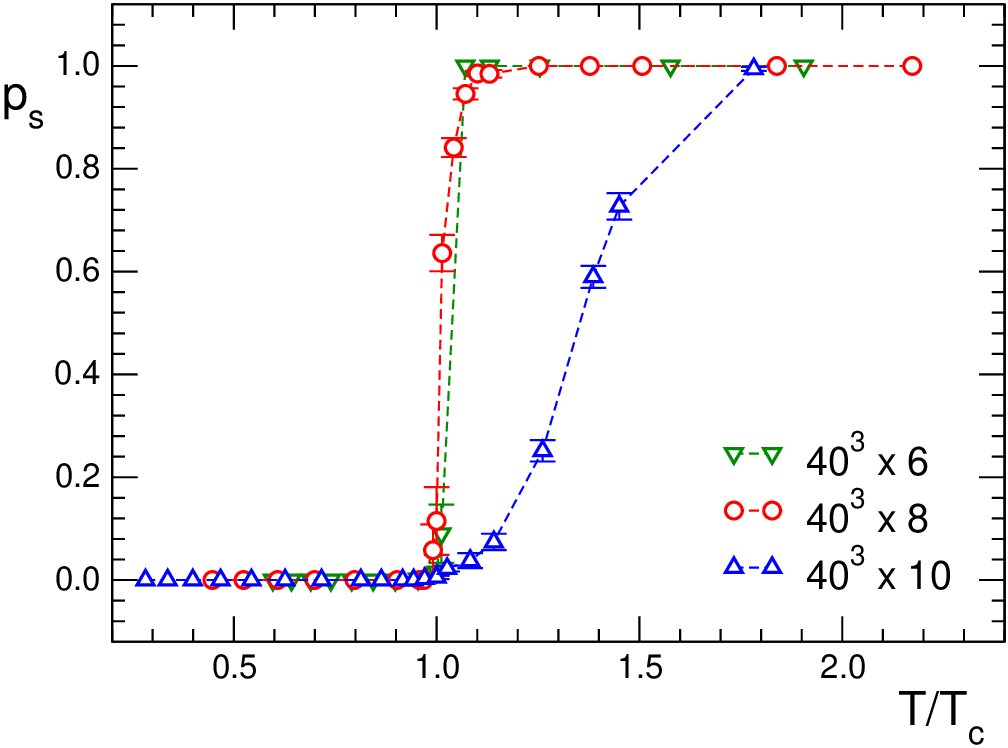}
\end{center}
\vspace{-5mm}
\caption{Probability for a spanning cluster as a function of temperature.  
\label{percprob}}
\end{figure}

The question of percolation can of course be addressed directly by looking at
the probability of having a spanning cluster. A spanning cluster is defined as a
cluster the extends from one end of the (spatial) lattice to the other. Here we
use periodic boundary conditions and thus we consider a cluster to be spanning
if all possible $x$-$y$ planes contain at least one site of the cluster. In
Fig.~\ref{percprob} we plot the average probability $p_{s}$ to have a spanning
cluster (i.e., the percolation probability) as a function of the temperature.
Again we compare the results for $N_t = 6, 8$ and $N_t = 10$ to assess the
approach to the continuum limit.

The probability $p_s$ to have a spanning cluster is essentially a step function
only for $N_t = 6$.  For $N_t = 8$ the probability increases somewhat slower,
and for $N_t = 10$ the percolation probability reaches 1 only at $T \sim 1.8 \,
T_c$. This finding  leads to the conclusion that when approaching the continuum
limit (increasing $N_t$) the onset of full percolation (probability for spanning
clusters reaches $p_s = 1$) does not coincide with the deconfinement transition at
$T_c$.

\begin{figure}[t]
\begin{center}
\hspace*{-13mm}
\includegraphics[width=10.0cm,clip]{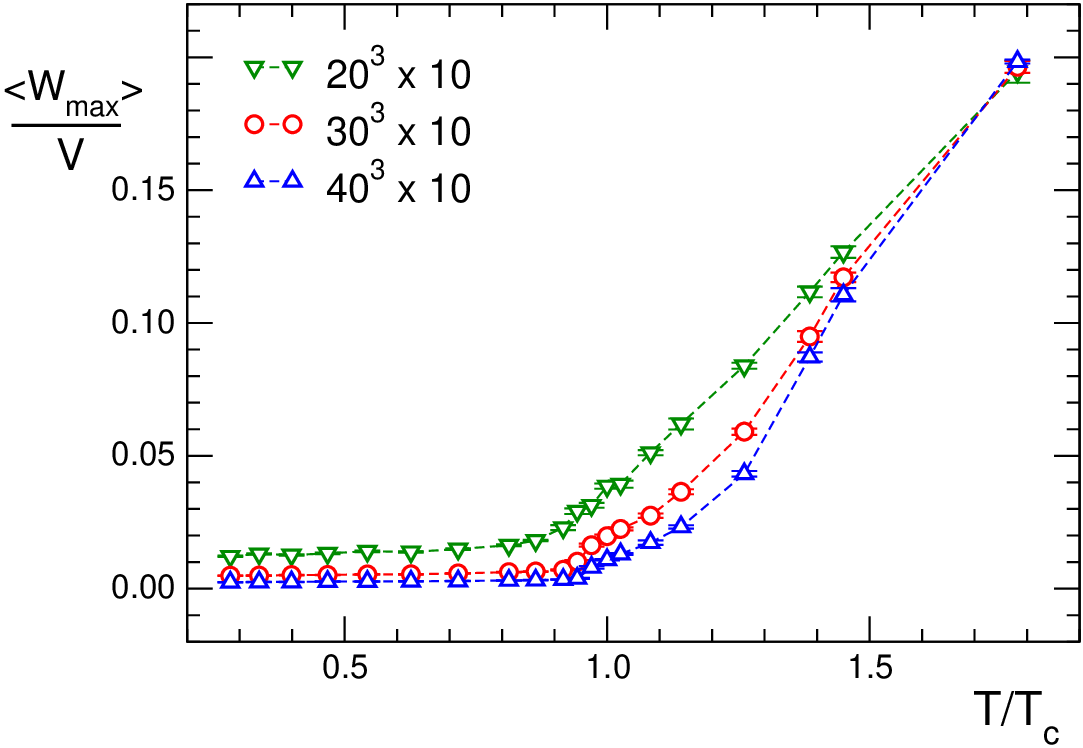}
\end{center}
\vspace{-5mm}
\caption{Finite volume study of the weight of the largest cluster plotted 
as a function of temperature.  
\label{finvol}}
\end{figure}

An important question is of course whether finite volume effects could be a
major effect: When increasing $N_t$ one has to decrease $a$ in order to keep the
temperature fixed. This decreasing lattice spacing $a$ then of course also
shrinks the spatial extent $L = a \, N_s$ in physical units. 
In order to assess such possible
finite size effects we compared our results for $20^3 \times 10$, $30^3 \times 10$
and $40^3 \times 10$. An example of the finite volume analysis is given in
Fig.~\ref{finvol} where we show the weight of the largest cluster as a function of 
the temperature, comparing three different spatial volumes. One finds that 
with increasing volume, the weight of the largest cluster even decreases slightly. This
decrease can be understand from the fact, that a finite hypertorus allows for additional
connections of sites around the periodic boundary conditions. This is a finite volume
effect that diminishes with increasing $N_s$ giving rise to the decreasing values of 
$W_{max}/V$.

The same finite volume analysis was performed also for the probability of  spanning
clusters $p_s$ and we found that the corresponding curves  vary only slightly, although
the volume changes by a factor of $8$. In particular we observed that the temperature
where $p_s$ reaches $p_s = 0.5$ is at $T = 1.35 \,  T_c$ for all three spatial volumes. We
thus confirm that the discrepancy between the deconfinement temperature and the point
where percolation is established is not a finite size effect.

Furthermore we have analyzed additional observable such as the gyration radius of the
clusters and again found that as one approaches the continuum limit there is no
coincidence of the deconfinement temperature with an onset of percolation
\cite{michael}.  

\vskip7mm
\noindent  
{\Large Discussion of the results}  

\vskip3mm
\noindent
In this paper we have analyzed the percolation properties of center clusters in pure SU($4$)
lattice gauge theory. To construct the center clusters we consider the phases of the
local Polyakov loops and use them to identify the nearest center element at each site of the
spatial lattice. Neighboring sites with the same center element are then assigned to the same
cluster. We studied various properties of the center clusters, and here in particular discuss our
results for the weight of the largest cluster and the probability to find a spanning cluster. 
An assessment of these observables on lattices with different lattice spacing shows that in the
continuum limit the deconfinement transition of SU($4$) lattice gauge theory 
does not coincide with an onset of percolation for
the center clusters. Cross checks with different spatial volumes rule out that finite volume
effects play a major role.

The finding we present here for SU($4$) is different from what was established for
SU($2$) and SU($3$)  pure lattice gauge theory \cite{satz,fortunato,gattringer}, where
indeed center clusters may be constructed such that the temperature for the  onset of
percolation agrees with the deconfinement temperature. For these two cases a more general
cluster construction is possible which does not automatically  link all neighbors with
equal center elements, but puts them in the same cluster only after some cutoff is
applied \cite{gattringer}, a step which corresponds to the construction of
Fortuin-Kasteleyn clusters that are known to percolate at the same temperature where 
Potts models with continuous transitions demagnetize \cite{fkclusters}. The construction
allows one to use the cutoff parameter of the clusters to establish a continuum limit
for the percolation description. For SU($4$) no such cutoff can be introduced as it would
further weaken the percolation properties of the clusters.

For gauge groups SU($N$) with values of $N$ even larger than $N = 4$, the clusters are thinned
out further: The number of center elements equals $N$ and if they are occupied with equal
probability each site is in a given sector with probability $1/N$. With increasing $N$ this
probability decreases and the clusters are thinned out and shrink. We conjecture that the 
center percolation picture of deconfinement fails for all $N$ larger than 3. Only the onset
of very strong correlations at $T_c$ 
between local Polyakov loops with phases near the same center element could give rise to
percolation. Our results show that such correlations are not strong enough to enable percolation
at $T_c$ for SU($4$), and we believe that the even stronger correlations necessary for  
higher SU($N$) (where the probability $1/N$ is even smaller) do not exist.

One may of course speculate in what respects the groups SU($2$) and SU($3$) are different. We 
point out that the effective theories \cite{znbreaking} for the center degrees of freedom, 
which govern the deconfinement transition of pure gauge theories are different for SU($2$) and
SU($3$). In general the corresponding center symmetrical effective action has the form \cite{znbreaking}
\begin{equation}
S \; = \; - \beta \sum_{<x,y>} \, \Big[ \, s_x \, s_y^\star \, + \, c.c. \, \Big] \; ,
\label{effact}
\end{equation}
where the sum runs over all nearest neighbors and $s_x$ is an element of the center group
$\mathds{Z}_N$, i.e., it is a phase $s_x = \exp( i 2 \pi k_x / N)$ with $k_x = 0,1, \, ... \, N-1$.
For $N = 2$ and $3$ it is possible to rewrite the action to the form
\begin{equation}
S \; = \; - \beta \, \sum_{<x,y>} \, A \, \delta_{k_x,k_y} \; + \; C \; ,
\label{potts}
\end{equation}
where $A$ and $B$ are trivial constants  and $\delta_{k_x,k_y}$ is the Kronecker delta.
In other words, for SU($2$) and SU($3$) the effective theory is a 2-state (3-state) Potts
model, where the agreement of the percolation of Fortuin-Kasteleyn clusters with the
demagnetization transition is established (exactly for the 2-state case
\cite{fkclusters}, numerically for the 3-state model \cite{first2}).  For $N > 3$ the
effective action (\ref{effact}) cannot be cast into the form (\ref{potts}) of the Potts
model. The reason is that for $N > 3$ the contribution of two neighboring spins assumes
more than two different values in the action  (\ref{potts}). The fact that the effective
theories for $N > 3$ have a form which is different from the $N = 2$ and $N = 3$ cases
might explain why  the center percolation description of deconfinement is not possible
for $N > 3$.

A second reason might be that for the weak first order transition of SU($3$) the
percolation  description is still applicable, while for the stronger first order
transition of SU($4$)  it fails. As already pointed out in the introduction, for
first order transitions the understanding of a possible relation between
percolation and thermal transitions is considerably less developed, and more
investigations are necessary. This work provides a particular counter example.

Finally we remark that another difficulty for a straight-forward interpretation  of the
results might be  that SU($4$) is a rank three group. As a consequence, as noted above, 
there is no simple effective Polyakov loop model for it. Indeed, any  such model contains
Polyakop loops in the three fundamental  representations, the $4$, ${\bar 4}$, and the
$6$. Depending on the relative  weight, the effective theory can be any type of model
from a  4-state Potts model to  two non-interacting Ising models
\cite{Strodthoff:2010dz,Wozar:2006fi}.  This would correspond to regarding the center
either as $\mathds{Z}_4$ or as  $\mathds{Z}_2\times \mathds{Z}_2$.  Our investigation was
assuming a $\mathds{Z}_4$-like  behavior,  but if this would not be appropriate, one
could speculate that a  different definition of center clusters could still show a
percolation  behavior at the deconfinement transition. However, note that in three
dimensions, the SU($4$) theory is  possibly close to the $\mathds{Z}_4$ option considered
here  \cite{Strodthoff:2010dz}. This possibility does not exists for SU($2$) and 
SU($3$), which harbor only a single way of representing the center.

\vskip5mm
\noindent
{\bf Acknowledgments:} The authors thank Hubert Antlinger, Szabolcs Borsanyi, 
Mike Creutz, Julia Danzer, Christian Lang and Alexander Schmidt
for valuable comments. 
The numerical calculations were done at the UNI-IT
clusters of the University Graz. A.~M.~was supported by the FWF under grant number M1099-N16 and by the 
DFG under grant number MA 3935/5-1.

\clearpage

\end{document}